\documentstyle[11pt]{article}
\begin{document}
\baselineskip=4.5mm
\newcommand{\be} {\begin{equation}}
\newcommand{\ee} {\end{equation}}
\newcommand{\Be} {\begin{eqnarray}}
\newcommand{\Ee} {\end{eqnarray}}
\def\lg{\langle}
\def\rg{\rangle}

\def\a{\alpha}
\def\b{\beta}
\def\g{\gamma}
\def\G{\Gamma}
\def\d{\delta}
\def\D{\Delta}
\def\e{\epsilon}
\def\k{\kappa}
\def\l{\lambda}
\def\om{\omega}
\def\Om{\Omega}
\def\t{\tau}
\noindent
\begin{center}
{\Large
{\bf
Response-theory for nonresonant hole burning: Stochastic dynamics
}}

\vspace{1cm}

\noindent
{\bf Gregor Diezemann} \\

Institut f\"ur Physikalische Chemie, Universit\"at Mainz,
Welderweg 15, 55099 Mainz, FRG
\\
\end{center}
\vspace{1cm}
\noindent
{\it
The time signals relevant for nonresonant spectral hole burning, a
pump-wait-probe method designed to investigate slow relaxation, are calculated.
The step-response function following the application of a high amplitude ac
field (pump) and an intermediate waiting period is shown to be the sum of the
equilibrium integrated response and a modification due to the preparation
via ac irradiation.
Both components are calculated for a class of stochastic dipole reorientation
models.
The results indicate that the method can be used for a clearcut
distinction of homogeneously and heterogeneously
broadened susceptibilities as they occur in the slow primary relaxation of
supercooled liquids and other disordered materials. This is because only in the
heterogeneous case is a frequency selective modification of the response
possible.
}

\vspace{0.5cm}

\vspace{1cm}
Disordered materials such as glasses, spin-glasses, disordered crystals
and proteins exhibit non-exponential relaxation behavior on macroscopic
time scales\cite{het.dyn}.
In the past decade several experimental techniques have been invented in order
to investigate the detailed nature of the non-exponential primary response
particularly of amorphous polymers and supercooled
liquids\cite{het.dyn},\cite{SRS91},\cite{CE95}.
These techniques allow to specifically select a (slow) sub-ensemble and
afterwards monitor its relaxation. They have shown the existence of
{\it dynamic} heterogeneities which, however, are not static
but fluctuate in time\cite{4dnmr}.
Recently the technique of nonresonant spectral hole burning (NHB)
has been developed and applied to study the primary relaxation of
supercooled liquids on the time scale of seconds\cite{nhb.sci}.
In the meantime NHB has also been used to investigate the slow
relaxation of disordered crystals\cite{KSB98}, amorphous
ion-conductors\cite{RB99} and spin glasses\cite{ralph99}.
The interpretation of the results has mainly been guided by the fact
that via the application of a large amplitude ac field of frequency $\Om$
($10^{-2}$Hz$\cdots 10^2$Hz) the sample absorbs energy of an amount
proportional to the imaginary part of the susceptibility evaluated at the
pump frequency $\Om$\cite{kubo}.
It was assumed that a frequency selective modification of the spectrum should
be feasible only if the response is given by a heterogeneous superposition of
entities relaxing at different rates.
However, the results of a recent numerical investigation of the
application of NHB to a solvable glass-model\cite{leticia00} suggest that
this view may have to be revised.
Beyond phenomenological interpretations a deeper unterstanding of a
pump-wait-probe experiment like NHB can be expected from a thorough analytical
analysis, which, however, has not been presented up to date.

It is the purpose of this Letter to provide a sound response theory for NHB.
The experimental protocol of NHB consists of the following
procedure\cite{nhb.sci}.
First, a large amplitude ac pump field $E_P\sin{(\Om \tau)}$
is applied to a sample in thermal equilibrium for a time
$t_p=n(2\pi/\Om), n=1,2,3,\cdots$.
After this pump period the system is left in zero field for a waiting time
$t_w$ before the step response is monitored via application of a
small amplitude dc field $E_S$ at time $t=0$.

In contrast to ordinary nonlinear response theory, I calculate the linear
response of a variable $S(t)$ to the static field $E_S$ {\it but} with a
disturbed initial state of the system prior to the application of $E_S$.
This initial state is calculated in order ${\cal O}(E_P^2)$.
Correspondingly, the response reads as
$R^*(E_S,E_P,t,t_w,t_p) = R(E_S,t) + \D R(E_S,E_P,t,t_w,t_p)$
where $R(E_S,t)$ denotes the equilibrium integrated response.
The 'modification' $\D R(E_S,E_P,t,t_w,t_p)$ originates from
the deviations from thermal equilibrium at time $t=0$ and consists of two
terms of order ${\cal O}(E_P)$ and ${\cal O}(E_P^2)$, respectively.
A subtraction of the signals following the application of a positive and a
negative dc field $E_S$ along with the addition of the signals obtained for a
positive and a negative pump allow to extract the relevant terms in
${\cal O}(E_S), {\cal O}(E_P^2)$. This corresponds to the phase cycle employed
in experiments\cite{nhb.sci}. The final result reads as:
\be\label{R*.pc}
R^*(t,t_w,t_p) = R(t) + \D R(t,t_w,t_p)
\ee
with $R(t)=\frac{1}{2}(R(E_S,t)-R(-E_S,t))\equiv R(E_S,t)$ and
\Be\label{dR.res}
\D R(t,t_w,t_p) = {\cal N}\!\!
\int_0^{t}\!d\tau {\rm Tr}\left\{ S(0)
e^{-i{\cal L}(t-\tau)}
[-i{\cal L}_1(E_S)]
e^{-i{\cal L}(\tau+t_w)}\D\rho_2(t_p)\right\}
\Ee
Here, ${\cal L}$ is the Liouvillian, ${\cal L}_1(E_S)$ the first order
perturbation due to the dc field and ${\cal N}$ is a constant.
$\D\rho_2(t_p)$ denotes the ${\cal O}(E_P^2)$-deviation from thermal
equilibrium of the density matrix directly after the pump. This results from
two terms, ${\cal L}_1(E_P)$ and ${\cal L}_2(E_P)$ of respective order
${\cal O}(E_P)$ and ${\cal O}(E_P^2)$\cite{FS90}.
It is evident from Eq.(\ref{dR.res}) that a non-vanishing $\D R(t,t_w,t_p)$
results only because the system has been driven out of equilibrium prior to the
linear response experiment and Eq.(\ref{R*.pc}) shows that the modified
response is just the sum of the ordinary linear response and the modification.
Therefore, in the NHB experiments the linear response starting from a modified
inital state (in ${\cal O}(E_P^2)$) is monitored as opposed to ordinary
nonlinear (dielectric) experiments.

In order to be able to further discuss the implications of
Eqns.(\ref{R*.pc},\ref{dR.res}), in the following I will consider explicit
models of stochastic dynamics without inertial terms. This is a reasonable
assumption for the slow relaxation processes of interest in the present
context.
The results obtained so far can be used directly for the corresponding
Fokker-Planck (FP) or master equation (ME) obeyed by the conditional
probability $P({\bf x},t)$, where ${\bf x}(t)$ denotes the
stochastic process under consideration\cite{vkamp81},\cite{risken}.
For a FP equation, the nonlinear response theory can be formulated in general
terms\cite{morita86}. In this case the resulting expressions are particularly
simple because higher order terms like ${\cal L}_2(E)$ vanish.

If a ME is considered instead the situation can be quite different.
The procedure of calculating the response as for a FP equation can be utilized
only if the transition probabilities in the ME are chosen in such a way that a
Kramers-Moyal expansion\cite{vkamp81} is possible.
If, however, the 'jump length' in the transition probabilities is large, the
question how to couple the transition probabilities to the field arises.
This can be understood most easily with the following argument. Denote the
transition probability for a $x_i\to x_k$ transition without externally
applied fields by $\k_{k,i}(t)=\k(x_k,x_i;t)$, the time dependent probability
distribution $p_i(t)=P(x_i,t)$ and the equilibrium probability by
$p_i^{eq}=p_i(t\to\infty)$ in a discrete notation.
From ordinary statistical mechanics the change in the
$p_i^{eq}\propto e^{-\b\e_i}$ with $\b=1/(k_BT)$
($k_B$ is the Boltzmann constant) and $\e_i$ the energy, is known.
If the field $E$ couples to a function $M({\bf x})$, these are just given
by $p_i^{eq}(E)=p_i^{eq}\exp{(\b M(x_i)E)}$.
How to change the $\k_{k,i}(t)$ when an external field is applied is however
not easily determined in the general case.
The only restriction is that detailed balance has to be obeyed and this will
be used in all what follows.
The most general form for the $\k_{i,k}(E,t)$ therefore is
\be\label{k.ik.E}
\k_{i,k}(E,t)=\k_{i,k}(t)e^{\b E\Psi_{i,k}}
	\quad\mbox{where}\quad \Psi_{i,k}=\a M(x_i)-(1-\a)M(x_k)
\ee
with $\a$ denoting a real number.
The perturbation series for the ME follows from  Eq.(\ref{dR.res})
if one substitutes $(-i{\cal L})$ by the master operator ${\cal W}$ and the
perturbation $(-i{\cal L}_n)$ by a corresponding ${\cal V}^{(n)}$ in
${\cal O}(E^n)$.
The matrix elements of ${\cal W}$ and ${\cal V}^{(n)}$ are given by:
\Be\label{W.V.ik}
&&({\cal W})_{i,k}=\k_{i,k}(t)-\d_{i,k}\sum_l\k_{l,k}(t)
\nonumber\\
&&({\cal V}^{(n)})_{i,k}=\frac{1}{n!}\left[\b E(t)\right]^n
    \left\{ \k_{i,k}(t)\Psi_{i,k}-\d_{i,k}\sum_l\k_{l,k}(t)\Psi_{l,k}\right\}
\Ee
The constant $\a$ is determined by calculating the long time limit of the
$p_k(E,t)$ as these have to coincide with $p_k^{eq}(E)$ in the presence of
external fields.
For example, for a ME that obeys a Kramers-Moyal expansion, one has
$\a=1/2$\cite{vkamp81}.

To fix the notation, in the following I will consider the reorientations of
rigid molecules with a permanent dipole moment $\mu$.
Thus $S(t)$ in Eq.(\ref{dR.res}) is identified with $\mu(t)$, the stochastic
process ${\bf x}(t)$ with the orientation $w(t)=(\phi(t),\theta(t),\psi(t))$
in terms of Euler angles and $M(w_i)=\mu_i=\mu\cos{\theta_i}$.
For simplicity, correlations due to the dipolar coupling of distinct
molecules and local field corrections are neglected throughout.
%
A simple way to model molecular reorientations is to assume that they
occur in a completely random way. This is usually termed as a 'random jump
model'\cite{DS99}.
In this model the transition probabilites for a change in orientation from
$w_i\to w_k$ are independent of $w_i$, $w_k$ and are given by
$\k_{k,i}=\G/N$ $\forall i,k=1,\cdots,N$. (If reorientations on a
sphere are considered, $N=8\pi^2$.)
The equilibrium probabilities are $p_i^{eq}=1/N$ and therefore the choice
of $\a$ is not determined a priori.

In the following, I will consider a slight modification of this random jump
model, for which the corresponding ME can still easily be solved analytically.
This model is defined by the choice
\be\label{kap.ik}
\k_{k,i}(t)=Zp_k^{eq}\G(t)\quad\mbox{with}\quad Z=\sum_i p^{eq}_i
\ee
which depends only on the final orientation of the transition.
From this fact it is evident that for this model $\a=1$.
Additionally, a time dependent rate $\G(t)$ is considered because this will
be used in the later discussion.
A simple choice yielding a non-exponentially decaying response is
\be\label{G.t.def}
\G(t)=b t^{b-1}/\t\quad b\in[0,1]
\ee
which reduces to a time independent rate $\t^{-1}$ for $b=1$.
Though the considered model is very simple minded, it still shows the
characteristic features of any stochastic model. In models with more complex
(time independent) $\k_{k,i}$ several rates $\l_m$ will occur as the
eigenvalues of the master operator ${\cal W}$. The mean decay rate in such
more sophisticated models can be defined by
$\frac{1}{N}\sum_m\l_m=\frac{1}{N}\sum_{i,k}\k_{i,k}$.
As this just equals $\t^{-1}$ for $b=1$, cf. Eq.(\ref{G.t.def}), the simple
model on average also reveals the features of more realistic ones.
Furthermore, if only two orientations are considered, $N=2$, the time honoured
double well potential (DWP) model with an asymmetry $(\e_1-\e_2)$ is recovered.

The response for the model defined by Eqns.(\ref{kap.ik},\ref{G.t.def}) is now
calculated according to Eqns.(\ref{R*.pc},\ref{dR.res}) using (\ref{W.V.ik}).
The resulting normalized response is given by $\Phi(t)=e^{-t^b/t}$ and
\be\label{d.Phi}
\D\Phi(t,t_w,t_p)=
-\frac{(\b E_P)^2}{\lg\D\mu^2\rg} A(\Om,b) (t^b/\t) e^{-t^b/\t}e^{-t_w^b/\t}
\ee
with
$\lg\D\mu^2\rg=\lg\mu^2\rg-\lg\mu\rg^2$, $\lg\mu^n\rg=\sum_i p_i^{eq}\mu_i^n$.
The magnitude of the modification at a given pump frequency, i.e.
the 'excitation profile', is given by ($t_p\equiv2n\pi/\Om$)
\Be\label{A.om.b}
A(\Om,b)=&&\hspace{-0.6cm}
\lg\mu\rg^2\lg\D\mu^2\rg(b/\t)^2e^{-t_p^b/\t}\!\!
\int_0^{t_p}\!\! dt_1\int_0^{t_1}\!\!dt_2
\sin{(\Om t_1)}t_1^{b-1}\sin{(\Om t_2)}t_2^{b-1}e^{t_2^b/\t}
\nonumber\\
&&\hspace{-0.6cm}
+\frac{1}{2}\left(\lg\mu\rg^2\lg\mu^2\rg-\lg\mu\rg\lg\mu^3\rg\right)
(b/\t)e^{-t_p^b/\t}\!\!
\int_0^{t_p}\!\! dt_1\sin{(\Om t_1)}^2t_1^{b-1}e^{t_1^b/\t}
\Ee
Eq.(\ref{A.om.b}) reveals one of the central results of the present
calculations, namely the frequency selectivity of NHB.
$A(\Om,b)$ vanishes for large and small burn frequencies $\Om$.
This directly implies that a non-vanishing modification of the response
can only be achieved for $\Om$ on the order of $\t^{-1}$.
The sign of $A(\Om,b)$ is determined by the prefactor of the second term
and therefore depends on the details of the assumptions for the $p_i^{eq}$.
Note that for a vanishing $\lg\mu\rg=0$ one has $A(\Om,b)=0$.
Furthermore, for a DWP model, it is easily seen that
$\lg\mu\rg^2\lg\mu^2\rg=\lg\mu\rg\lg\mu^3\rg$ and the second terms of
$A(\Om,b)$ vanishes.
The same holds approximately, if only small asymmetries are assumed in the
general case. Therefore, in the following calculations the second term of
$A(\Om,b)$ will be neglected throughout.

A transparent expression allowing a simple interpretation for the excitation
profile is obtained for the special case $b=1$:
\be\label{A.om}
A(\Om,1)\simeq
\frac{3}{2}\lg\mu\rg^2\lg\D\mu^2\rg\e''(\Om\t)\e''(2\Om\t)(1-e^{-t_p/\t})
\ee
Here, $\e''(m\Om\t)=m\Om\tau/(1+m^2\Om^2\tau^2)$.
Therefore, $A(\Om,1)$ is determined by the absorbed energy and higher order
terms, in accord with what is expected on physical grounds\cite{nhb.sci}.

In Fig.1 the normalized excitation profile $A(\Om,b)/A_{max}$ is
plotted as a function of the pump frequency for $b=1$ and $b=0.7$ and one pump
cycle.
For $b=1$, the excitation profile is considerably narrower than for $b=0.7$
and also than the Debye susceptibility $\e''(\Om\t)$.
This means that the selection is restricted to a narrow band of frequencies.
Decreasing $b$ gives rise to broader excitation profiles.

Important features of the modification $\D\Phi(t,t_w,t_p)$ given in
Eq.(\ref{d.Phi}) are the following.
(i) If $\D\Phi(t,t_w,t_p)<0$, the modified response $\Phi^*$ decays faster
than the equilibrium response. In the approximation used in Eq.(\ref{A.om})
this always holds.
The fact that $\D\Phi(t,t_w,t_p)\neq 0$ only for $\Om$ on the order of
$\t^{-1}$ has already been discussed above.
(ii) $\D\Phi(t,t_w,t_p)$ is only non-zero in a finite intervall of
time $t$, determined by $b$. The regime of $\D\Phi(t,t_w,t_p)\neq 0$ becomes
larger with decreasing $b$ (i.e. the 'spectral holes' become broader).
The maximum value of the modification is found at the time
$t_{max}=(1/\t)^{1/b}$.
For a homogeneous response (a fixed value of $\t$) a variation of the pump
frequency will only alter the overall amplitude of the modification.
Therefore, it is not possible to perform a frequency selective modification in
this case.
(iii) The modification dereases as a function of the waiting time $t_w$.
This re-equilibration proceeds with the same relaxation time $\t$.
The reason for this behavior becomes evident from the form of the modified
orientational distribution after the pump,
$p^{\rm Mod}_k(t_p)\sim (\mu_k-\lg\mu\rg)p_k^{eq}$, which in turn relaxes
with the dipole relaxation time $\t$.
For more sophisticated models, other relaxation times $\l_m^{-1}$ may occur
here.
The important finding is that there is no extra time scale for
re-equilibration.
It should be mentioned that similar results are found for a variety of
stochastic models of dipole-reorientation\cite{greg.nhb}.

In order to further clarify the implications of the calculations, in the
following I will consider specific examples.
Often a stretched exponential function is used to parametrize the equilibrium
response, $\Phi(t)=e^{-(t/\tau_K)^{\b_K}}$ with $\b_K$ smaller than
unity, see e.g.\cite{het.dyn}.
In a {\it heterogeneous} scenario, this response is viewed as
originating from a distribution of $\t$\cite{het.dyn}, i.e.
$\Phi(t)=\int d\t g(\t)e^{-t/\t}$ with some distribution function
$g(\t)$, whereas $\Phi(t)$ is assumed to be intrinsically non-exponential
in a {\it homogeneous} scenario.
It is instructive to compare directly the results of calculations for the two
extreme scenarios for a realistically broadened susceptibility.
Such a comparison of $\D\Phi(t)$ is shown in Fig.2 for (a) a heterogeneous
and (b) a homogeneous scenario with $\Phi(t)=\exp{(-(t/1s)^{0.7})}$.
This means, I used $b=1$ and a corresponding distribution of $\t$ in the
calculations for case (a) and $b=0.7$, $\t^{-b}=1s$ in (b).
In the heterogeneous case it is evident that the time at which the maximum
modification shows up strongly depends on the burn frequency $\Om$. This
demonstrates the frequency selectivity of NHB.
Such a dependence is missing completely in the homogeneous case.
In the heterogeneous case the maximum modification shifts towards longer times
with decreasing $\Om$.
This is because those relaxation processes with a $\t$ yielding the
maximum $A(\Om,1)$ (cf. Eq.(\ref{A.om}) and Fig.1) contribute most to
$\D\Phi(t)$.
The functional dependence of $\D\Phi_{max}=\D\Phi(t_{max},t_w=0,t_p)$
upon $\Om$ depends sensitively on the choice of the relaxation time
distribution (e.g. on the value of $\b_K$).
Generally, there will be a rather strong dependence for large
$\Om$ which crosses over to a $\Om$-independence of $\D\Phi_{max}$ when
$1/\Om$ reaches the smallest relaxation time of the distribution.
Exchange processes, which may be responsible for fluctuations of relaxation
rates, may partially suppress the $\Om$-dependence\cite{greg.nhb}.

Also included Fig.2(a) are experimental data on propylene
carbonate\cite{nhb.sci}.
It is remarkable that all main features of the data can be described by
the simple heterogeneous model.
The position of the modifications and the relative amplitudes are in
quantitative agreement with the data. Also the width and the asymmetry are
described with high accuracy.
A similar quantitative agreement is obtained for the $t_w$-dependence of
experimental data on supercooled liquids\cite{greg.nhb}.

So far, I have considered stochastic models for dipole reorientations.
All these models show qualitatively the same features. In particular it is
found that the re-equilibration during the waiting time $t_w$ takes place on
the time scale of the intrinsic relaxation time ($\t$ in the above
examples).
In an experiment on a relaxor ferroelectric a re-equilibration time much
longer than $1/\Om$ has been found\cite{KSB98}.
In order to investigate the theoretical conditions for such an effect I
consider a simple 'nonlinear Debye model'.
A physical realization of this scenario may be a system where the
dynamics is dominated by domain wall depinning\cite{nat90}.
The polarization is assumed to relax with a correlation time
$\tau_0=\tau_\infty e^{\b U_a}$ where $U_a$ is the activation (pinning)
energy and  $1/\tau_\infty$ an attempt frequency.
Application of an electric field will change the activation energy roughly
by an amount $U_a(t)=U_a(0)-cP_{sp}E(t)$, where $c$ is a constant and
$P_{sp.}$ the spontaneous polarization\cite{GTB96}, which in the simplest
case is proportional the the applied field, $P_{sp.}\propto E(t)$.
The relaxation of the deviation from thermal equilibrium of $U_a(t)$ is
assumed to obey a linear law:
$\partial_t\d U_a(t)+\g\d U_a(t)=c\partial_t\left[E(t)^2\right]$
(meaningful for $\g<1/\tau_0$), where the rate $\g$ may depend on the
magnitude of deviation from equilibrium.
This expression together with the equation for P(t),
$\partial_t P(t) + (1/\tau)P(t)= \chi (1/\tau)E(t)$, can be solved in
${\cal O}(E_S)$ and ${\cal O}(E_P^2)$ for the NHB situation.
Here, $\chi$ denotes the dielectric susceptibility and
$\tau=\tau_0 e^{\d U_a(t)}$.
The calculation leads to results similar to those obtained above with
$\Phi(t)=e^{-t/\tau_0}$ and
$\D \Phi(t,t_w,t_p)\propto e^{-\g t_w}$.
The important point is that in this model $1/\g$, and {\it not}
$\tau_0$ is the relevant time scale for re-equilibration.

In conclusion, I have shown that NHB can be understood as a linear response
experiment starting from a nonlinearly perturbed initial state.
Generally, it is always possible to separate the linear response from the
effects of the pump process.
For systems with stochastic dynamics the method is clearly able to
discriminate between homogenously and heterogeneously broadened dielectric
spectra.
A frequency selective modification of the response is possible only in the
heterogeneous case.
The re-equilibration during the waiting time is determined by the relaxation
of the modified orientational distribution created during the pump period.
The hole recovery does not show the appearence of a second time scale.
A longer time scale for re-equilibration, if it is observed, has to be
attributed to intrinsic non-equilibrium effects.

\vspace{0.2cm}
I am grateful to R. B\"ohmer, H. Sillescu, O. Kircher, G. Hinze and
R. Schilling for very fruitful discussions and comments.

%

%
\subsection*{Figure captions}
\begin{description}
\item[Fig.1 : ] Normalized 'excitation profile' $A_n(\Om,b)=A(\Om,b)/A_{max}$
versus $\Om/\Om_{max}$ for $b=1,\Om_{max}\t=0.736$ (full line) and
$b=0.7,\Om_{max}\t=0.185$ (dashed line).
Also shown is the imaginary part of the susceptibility, $\e''(\Om\t)$,
where $\t$ is the relaxation time, (dotted line) for comparison.
The lines for $A_n(\Om,0.7)$ and $\e''(\Om\t)$ have been shifted by $0.5$ and
$1.0$ units, respectively.
\item[Fig.2 : ] $10^3\D\Phi(t)$ versus rescaled time $t /t_{max}$ for (a) a
heterogeneous and (b) a homogeneous scenario for various burn frequencies
$\Om$ and $\b^2E_P^2/\lg\D\mu^2\rg=0.05$.
In both cases the equilibrium response decays as
$\Phi(t)=e^{-(t/1s)^{0.7}}$.
The time $t_{max}=1.0$s in case (a) and $t_{max}=0.095$s in case (b).
The used frequencies are:
heterogeneous scenario: $\Om\t=5.0$ (1), $1.0$ (2), $0.2$ (3)
$0.1$ (4);
homogeneous scenario: $\Om\t=5.0$ (1), $1.0$ (2), $0.1$ (3), $0.05$ (4);
Experimental data in (a) are adapted from ref.\cite{nhb.sci}. Here,
$\Om\t$=1.02 (2) and $\Om\t$=0.203 (3).
\end{description}
\end{document}